\documentclass[aps,pra,floatfix,superscriptaddress,preprint, showpacs,groupedaddress]{revtex4-1}

\usepackage{graphicx}
\usepackage{graphics}
\usepackage{amssymb}
\usepackage{amsmath}
\usepackage{epsfig}
\usepackage{latexsym}
\usepackage{color}
\usepackage{rotating}
\usepackage{subfigure}
\usepackage{float}

\begin{document}

\title{Phonon number measurements using single photon opto-mechanics  }

\author{S. Basiri-Esfahani }
\email{Author to whom any correspondence should be addressed. Electronic mail: s.basiri@uq.edu.au}
\author{U. Akram}
\author{G . J. Milburn}

\affiliation{Centre for Engineered Quantum Systems, School of Mathematics and Physics, The University of Queensland, St Lucia, QLD 4072, Australia}
\begin{abstract}
We describe a system composed of two coupled optical cavity modes with a coupling modulated by a bulk mechanical resonator. In addition, one of the cavity modes is irreversibly coupled to a single photon source. Our scheme is an opto-mechanical realisation of the Jaynes-Cummings model where the qubit is a dual rail optical qubit while the bosonic degree of freedom is a matter degree of freedom realised as the bulk mechanical excitation. We show the possibility of engineering phonon number states of the mechanical oscillator in such a system by computing the conditional state of the mechanics after successive photon counting measurements.
\end{abstract}

\maketitle

\section{Introduction}
\label{intro}
Quantum opto-mechanics \cite{Vahala,Girvin,Aspelmeyer,WM} provides an exciting context in which to investigate the interaction of light and the collective motion of bulk mechanical systems in the quantum regime. More generally, quantum opto-mechanics provides a good example of an engineered quantum system: a meso/macroscopic device specifically engineered so that collective degrees of freedom can be subject to coherent quantum control and measurement. Recent achievement of the quantum ground state of a mechanical resonator in electro \cite{Cleland, Teufel1} and opto-mechanical systems \cite{Painter} has meant that we can now begin to observe and eventually manipulate quantum states \cite{Akram,Chen} at the mesoscopic scale. A recent opto-mechanical experiment towards this direction has been carried out illustrating an efficient quantum interface between optical photons and mechanical phonons \cite{Kippenberg}. An obvious next step is understanding and implementing single photon optical non linearities in such systems. Recent results for single photon opto-mechanics describe novel features such as the photon blockade effect \cite{Prabl, Kronwald} and cavity resonance shift \cite{He}, preparation of non gaussian mechanical states \cite{Nunnenkamp,Ludwig}, and mechanical superpositions \cite{Bouwmeester}.  

The model of this paper is based on two optical cavity modes with a coupling modulated by a mechanical degree of freedom. This can be realised in a number of different ways, but we will focus on the scheme of Chang et al. \cite{Chang}.   In this scheme a coherent exchange of photons between two photonic crystal defect cavities is modulated by a common mechanical degree of freedom. This implementation leads to a large coupling strength between the optical and mechanical modes and is a promising direction to achieve strong coupling at the single photon level.   Another implementation could be based on a single bulk flexural mode driven by the opposing radiation pressure forces of two optical cavity modes. If the cavity modes are coupled, transformation to normal modes leads to a model in which the normal mode coupling is modulated by the mechanical displacement \cite{Marquardt2012}.

In the present work we consider single photon driving of one unit in the opto-mechanical crystal of Chang et al.,  and show how this system may be mathematically described using the Jaynes-Cummings model \cite{JC} when the optical modes are excited by single photon states. A similar analogy to the Jaynes Cummings Hamiltonian has been studied previously in an atom assisted cavity optomechanical system in \cite{Yue Chang}. In our study the bosonic component of the Jaynes-Cummings model becomes the mechanical degree of freedom while the two-level component of the Jaynes-Cummings model is a dual-rail single photon qubit. The Jaynes-Cummings analogy can be exploited to realise  a measurement of the phonon number of the mechanical degree of freedom following a modified version of the measurement scheme of  Guerlin et al. \cite{Haroche}. Additionally, we condition the system on photon counts and calculate conditional states of the mechanical resonator. 

\begin{figure}[h!] 
   \centering
   \includegraphics[scale=0.6]{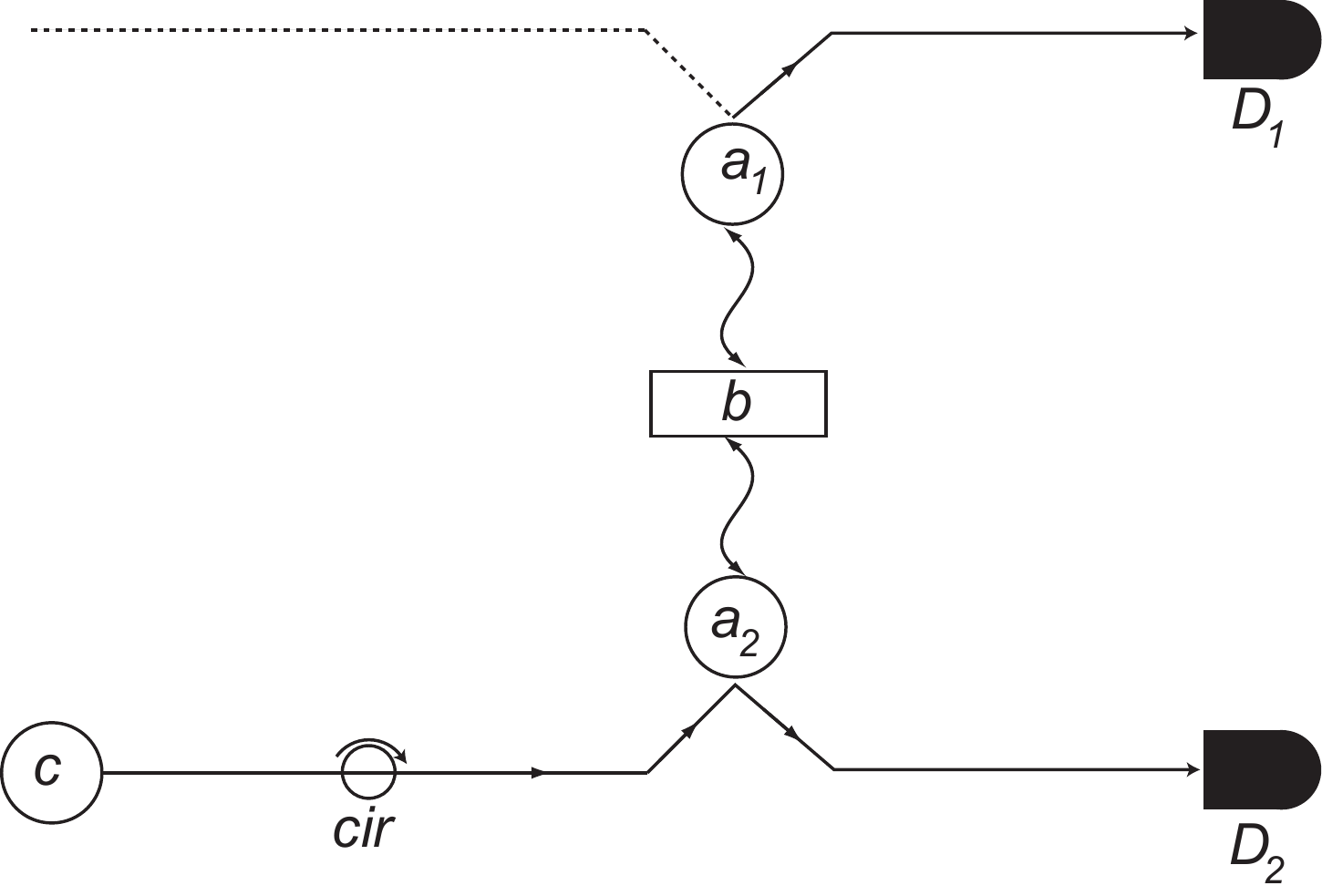} 
   \caption{A schemata for an opto-mechanical system in which two single photonic cavity modes $a_1, a_2$,  coherently exchange photons at a rate proportional to the displacement of a collective mechanical degree of freedom, $b$. The single photon source cavity is labelled $c$ and excites cavity $a_2$ irreversibly through the presence of a circulator, labelled $cir$. Single photon counters are labelled $D_1, D_2$.   }
   \label{scheme}
\end{figure}

A schematic diagram of the model we treat is given in figure \ref{scheme}. Two optical cavity modes are coupled in such a way that the coupling is proportional to a mechanical displacement.  This leads to a Raman process in which photons are exchanged between the cavities by absorbing or emitting a phonon.  The electrical field in each cavity is described by a single mode with photon annihilation and creation  operators $a_i, a_i^\dagger\  (i=1,2)$. The mechanical degree of freedom is described as single harmonic oscillator with phonon annihilation and creation operators $b,b^\dagger$. 
We will assume that one optical cavity, $a_2$,  is excited by emission from a single photon source modelled as another optical cavity (the source)  prepared with exactly one photon. The single photon excitation of $a_2$ is an irreversible process due to insertion of a circulator between the source and cavity $a_2$.  The injected photon can be reflected from cavity $a_2$ or absorbed and then remitted, to be detected at a single photo counter $D_2$. We also allow for the possibility  of  emission from cavity $a_1$ which can be monitored by detector $D_1$.  

The Hamiltonian for the opto-mechanical system, comprising the two cavity modes and the mechanical degree of freedom, may be written as \cite{Chang, Marquardt2012} 
\begin{equation}
H=\hbar\omega_1 a^\dagger_1 a_1+\hbar\omega_2 a_2^\dagger a_2 +\hbar\omega_m b^\dagger b+\hbar g(b+b^\dagger) (a_1^\dagger a_2+a_1 a_2^\dagger),
\end{equation}
where $\omega_{1/2}$ is the resonance frequency of each optical cavity, $\omega_m$ is the mechanical resonance frequency and $g$ is the rate of coherent coupling between the mechanics and the two optical cavities. Other realisations of such a three mode optomechanical interaction include \cite{Cheung} using the membrane in the middle model and also \cite{Zhao} in the context of an opto-acoustic parametric amplifier. 
 We now move to an interaction picture and assume that the system is so designed that  $\omega_2=\omega_1+\omega_m$. The opto-mechanical interaction picture Hamiltonian, including only resonant terms,  is 
\begin{equation}
\label{om-ham}
H_{om}=\hbar g(b^\dagger a_1^\dagger a_2+ba_1 a_2^\dagger).
\end{equation}
 This represents a kind of coherent Raman process whereby one photon from cavity $a_2$ is transferred into cavity $a_1$ simultaneously exciting one phonon in the mechanical degree of freedom. 
 
 The analysis is greatly simplified if we assume that, at most, there is one photon in the system at any time. In that case, the opto-mechanical interaction can be regarded as an interaction between a qubit and a simple harmonic oscillator with the qubit states defined as 
\begin{eqnarray}
|0\rangle & = & |1\rangle_1|0\rangle_2 ,\\
|1\rangle & = & |0\rangle_1|1\rangle_2 ,
\end{eqnarray}
where $|n\rangle_i$ are photon number eigenstates for cavity $a_i$.  This is the dual rail encoding used in linear optical quantum computing schemes \cite{KLM,Stannigel}.   On this restricted subspace we can define
\begin{eqnarray}
a_1^\dagger a_2  & = & |0\rangle\langle 1|\equiv   \sigma_- ,\\
a_1a_2^\dagger & = & |1\rangle\langle 0|\equiv  \sigma_+ ,
\end{eqnarray}
where $\sigma_{\pm}$ are the usual raising and lowering operators for a pseudo-spin system. The opto-mechanical interaction Hamiltonian can then be written in terms of the Jaynes-Cummings Hamiltonian 
\begin{equation}
\label{om-JC}
H_{om}=\hbar g(b\sigma_++b^\dagger \sigma_-).
\end{equation}

Each cavity mode is treated as a single sided cavity with photon  decay rates given by $\kappa_{1/2}$ for each cavity mode respectively.  The restriction to a single sided cavity is purely for simplicity in presenting our argument. It is straightforward to include loss at a second cavity mirror. The additional output port, if not monitored, provides an additional decay channel for the single photon.  Any photon lost through the unmonitored port is never detected and thus leads to an increase  in the number of trails required for a successful detection at the monitored port. 

 The single photon source, modelled as a source cavity prepared with a single photon \cite{Akram}, is irreversibly coupled into the input of the cavity $a_2$. The emission rate from the source cavity is $\gamma$. The input single photon states thus constitute a pulse with an exponential temporal profile with lifetime $\gamma^{-1}$. Other single photon excitations mechanism could be implemented including a quantum dot source embedded in the waveguide as in Schwagmann et al.  \cite{Shields}.   
 
 We model this kind of single photon excitation process using the theory of cascaded quantum systems \cite{Carmichael, Gardiner} to give a master equation for the dynamics of the OM system plus the source cavity
 \begin{equation}
 \label{OM-me}
\frac{d\rho}{dt}=-\frac{i}{\hbar}[H_{om},\rho]-\frac{\sqrt{\gamma\kappa_2}}{2}[ca_2^\dagger-c^\dagger a_2,\rho]+{\cal D}[J]\rho+\kappa_1{\cal D}[a_1]\rho+\gamma_m(\bar{N}+1){\cal D}[b]\rho+\gamma_m\bar{N}{\cal D}[b^\dagger]\rho ,
\end{equation}
were $c,c^\dagger$ are the annihilation and creation operators for photon number states in the source cavity, the superoperator ${\cal D}$ is defined by 
\begin{equation}
\label{superoperator}
{\cal D}[A]\rho =A\rho A^\dagger -\frac{1}{2}( A^\dagger A\rho +\rho A^\dagger A),
\end{equation}
 and the detection operator is $J=\sqrt{\gamma} c+\sqrt{\kappa_2} a_2$ which consists of the sum of two terms: the first term shows that the photon can be detected directly from the source before it is transmitted to cavity $a_2$ and the second term shows a photo-detection from within cavity $a_2$. We have also assumed that the source cavity is on resonance with the receiver cavity, $a_2$.   If cavity-2 had two output mirrors, we would include an additional decay term in the master equation of the form $\kappa'_2{\cal D}[a_2]\rho$ where $\kappa'_2$ is the loss rate through the second, unmonitored output mirror of this cavity. 
 
 The initial state is taken to be such that at $t=0$ there is one photon in the source and no photons in either cavity $a_1$ or cavity $a_2$, while the mechanical system is in an arbitrary coherent state $|\beta\rangle_b=e^{-|\beta|^2/2}\sum_n \beta^n/\sqrt{n!}|n\rangle_b$ where $b^\dagger b|n
\rangle_n =n|n\rangle_n$. The total initial state is thus
\begin{equation}
\label{initial}
|\Psi(0)\rangle = |1\rangle_c|0\rangle_1|0\rangle_2|\beta\rangle_b .
\end{equation}

Given this initial state, we now ask for the conditional state of the mechanics given that the photon is detected  at $D_2$ at a time $t_1$. In the ideal case we would like the photon to be detected with certainty at $D_2$, however in reality it could fail to be counted due to non unit quantum efficiency in the detector,  lost either through emission out of cavity $a_1$, scattered out of the input and output channels, or perhaps not emitted by the source at all. However assuming the ideal case, every photon emitted by the source is counted at $D_2$ or $D_1$. If we further assume  the decay rate of cavity $a_1$ is negligible (although we include it in our analysis), every photon emitted by the source is counted so that no information is lost and each photon detection event at $D_2$ gives a single bit of information. To gain additional information about the mechanics we can simply repeat the process, each time preparing a single photon in the source and using the conditional mechanical state obtained by the previous detection event. 

\section{Measurement of phonon number}
\subsection{Measurement and the Jaynes-Cummings model}
\label{simple-JC-model}
At first sight the interaction Hamiltonian in Eq.(\ref{om-JC}) does not look as if it could realise a measurement of the phonon number operator, $b^\dagger b$, for the mechanics. However given the Jaynes-Cummings representation in Eq.(\ref{om-JC}) we can use the results of the Haroche group \cite{Haroche-book} to see how it may be configured so as to yield information on the state of the mechanical system.  We will first review the simpler case in which interactions are deterministic. 

Let us suppose that we prepare the two-level  system in  the state $|1\rangle$ and that the interaction with the bosonic system proceeds for a time $\tau$ at which point we make an arbitrarily accurate measurement of the qubit state. The result of this measurement is a single binary number $x$.  The resulting conditional  (unnormalised) state for the bosonic degree of freedom is 
\begin{equation}
|\tilde{\psi}^{(x)}\rangle =E(x) |\psi(0)\rangle_b ,
\end{equation}
where
\begin{equation}
E(x)=\langle x|e^{-i\theta (b\sigma_++b^\dagger\sigma_-)}|1\rangle ,
\end{equation}
where $\theta=g\tau$. It is a simple matter to show that
\begin{eqnarray}
\label{E1}
E(1)  & = &  \cos(\theta\sqrt{bb^\dagger}) ,\\
\label{E2}
E(0)  & = & -ib^\dagger (bb^\dagger)^{-1/2}\sin(\theta \sqrt{bb^\dagger}) .
\end{eqnarray}
The probability to obtain the result $x$ is simply the required normalisation of the unnormalised conditional state 
\begin{equation}
p(x)=\langle \tilde{\psi}^{(x)}|\tilde{\psi}^{(x)}\rangle =\mbox{}_b\langle \psi(0)|E^\dagger (x)E(x)|\psi(0)\rangle_b .
\end{equation}
As $\sum_x  E^\dagger (x)E(x)=1$ this probability distribution is normalised. 

As the measurement operators in Eqs. (\ref{E1}) and (\ref{E2}) commute with the mechanical phonon number, this model describes a coarse-grained phonon number measurement; coarse-grained because there is only one bit of information per trial. It is not however a quantum nondemolition measurement (QND) as the interaction Hamiltonian between the probe and the mechanical system in Eq. (\ref{OM-me}) does not commute with the number operator. In fact, as we started with an interaction that is linear in the mechanical displacement we could not expect a QND measurement of the phonon number which requires an interaction that is at least quadratic in the displacement \cite{Adil}. It is possible to configure the system discussed in this paper in the strong dispersive regime so that there is an effective coupling to the displacement squared thus realising a QND phonon number measurement \cite{Marquardt2012}.

Suppose the initial state of the bosonic degree of freedom is a coherent state $|\psi(0)\rangle=|\beta\rangle_b$, where
\begin{equation}
|\beta\rangle_b=e^{-|\beta|^2/2}\sum_{n=0}^\infty \frac{\beta^n}{\sqrt{n!}}|n\rangle_b ,
\end{equation}
so that the initial number distribution is the Poisson distribution
\begin{equation}
P_0(n)=e^{-|\beta|^2} \frac{|\beta|^{2n}}{n!} .
\end{equation}
If the result of the measurement was $x=1$, the number distribution for the conditional state now becomes
\begin{equation}
P_1(n)=[p(1)]^{-1} \cos^2(\theta \sqrt{n+1})P_0(n) .
\end{equation}

\begin{figure}[h!]
 \centering
   \includegraphics{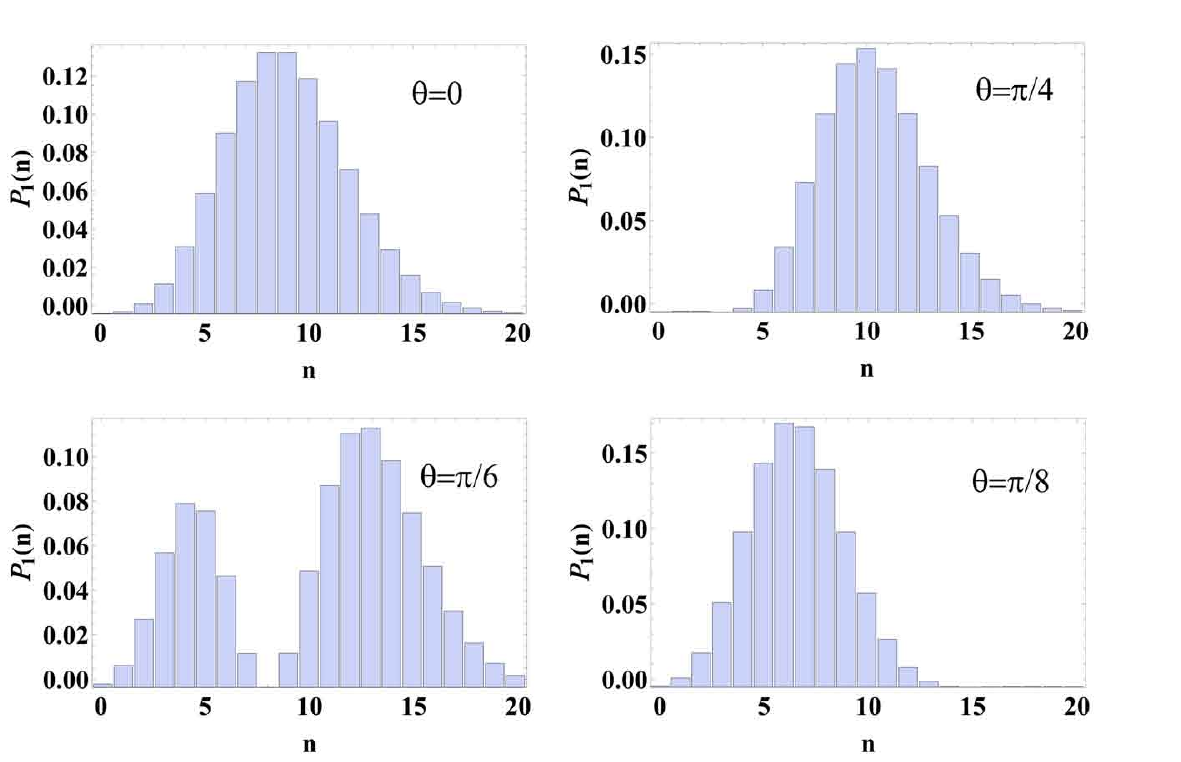} 
   \caption{The conditional number distribution after one readout with $x=1$ for various values of $\theta$. In all cases $\beta=3$. }
   \label{fig2}
\end{figure}

\begin{figure}[h!]
   \centering
   \includegraphics{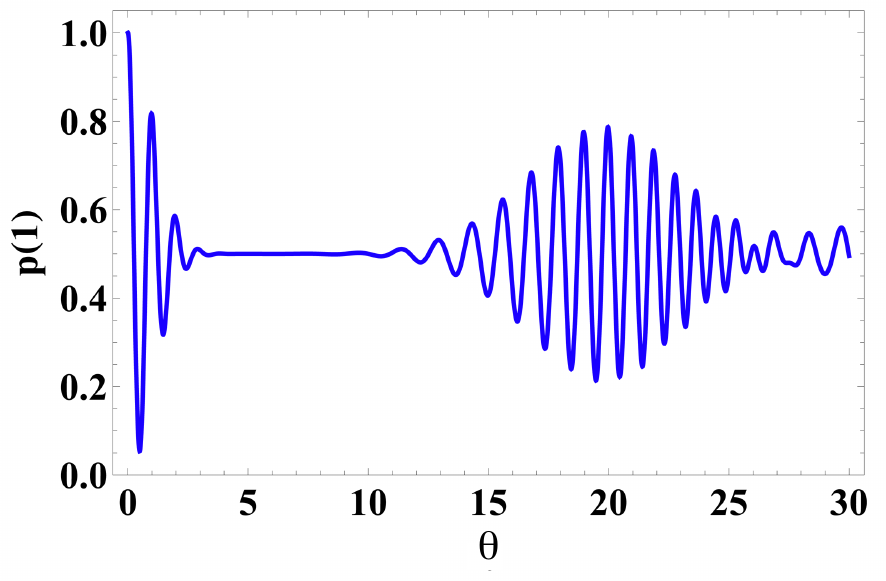} 
   \caption{The probability to obtain the result $x=1$ for $\beta=3$ versus $\theta=g\tau$. }
   \label{fig3}
\end{figure}

In figure \ref{fig2}  we plot this distribution, together with the initial distribution,  for various values of the parameter $\theta$. The important feature to note is that for particular values of $\theta$, in this case $\theta=\pi/6$, the distribution is very different from the initial poisson distribution.  Figure \ref{fig3} shows the probability for this outcome s a function of $\theta$, $p(1)=\langle \cos(\theta\sqrt{n+1})\rangle$ with the average taken over the number distribution prior to measurement. Clearly this average is bounded by one and, from the figure, we see that it oscillates around $1/2$. This indicates that the information provided in a single measurement is at most one bit. 

\begin{figure}[h!]
   \centering
   \includegraphics{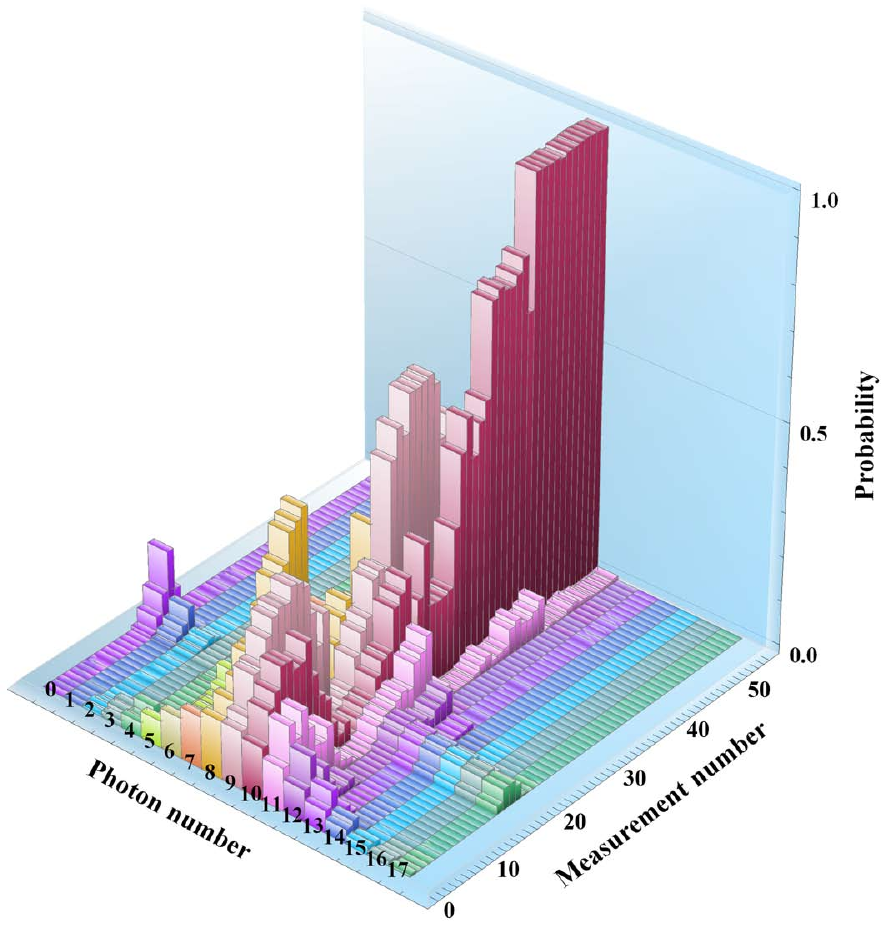} 
   \caption{Photon number probabilities versus photon number and number of readout, all with $x=1$, $\beta=3$ and for various values of $\theta$. }
   \label{fig4}
\end{figure}

We now consider what happens if we repeat the measurement using the conditional state from the first measurement as the initial state for the bosonic degree of freedom for the next measurement, but change the value of $\theta$.   In figure \ref{fig4} we plot the conditional number distribution for fifty five measurements that all gave the result $x=1$, but with different values of $\theta$.
We see that after the sequence of measurements, for appropriate values of $\theta$, the conditional state can approach a number state. 

To get an idea of how many measurements are required to reach a number state, we can consider the case for which the value of $\theta$ in each trial is the same. In $N$ such trials, the final distribution is given by
\begin{equation}
P_N(n)=[p(1,1\ldots, 1)]^{-1} \cos^{2N}(\theta \sqrt{n+1})P_0(n) .
\end{equation}
where the normalisation is simply the probability for this history of measurement results, $p(1,1\ldots, 1)=\sum_{n=0}^\infty \cos^{2N}(\theta \sqrt{n+1})P_0(n)$. The cosine factor is a periodic comb-like distribution with respect to $n$. If we choose $\theta$ to align the comb so that a peak is on the mean phonon  number, we can estimate the width of the distribution $P_N(n)$. We thus choose $\theta\sqrt{\bar{n}}=\pi$, and expand the cosine factor around this value to find that the distribution can be approximated by a Gaussian with variance 
\begin{equation}
W\approx \frac{2\bar{n}{^2}}{\pi N}
\end{equation}
This indicates that the number of measurements we would need to make to approach  a Fock state, for which $W=1$,  scales as $N\propto \bar{n}^2$.  This suggests that one should prepare the mechanical resonator in a state with a small value of $\bar{n}$ in order to have a reasonable chance of getting to a Fock state before mechanical heating takes over. 

 In the Haroche experiments \cite{Haroche-book} the values of $\theta=g\tau$  are kept approximately  constant from trial to trial by adjusting the velocity distribution of the atoms passing through the cavity. In the model of this paper however the interaction times $\tau$ are stochastic due to the probabilistic  character of the photon dwell time in the cavity. However it is clear that the interaction times are most likely to be of the order of $\kappa_2^{-1}$ so that the condition $\theta\sqrt{\bar{n}}=\pi$ translates to $g/\kappa \sim \pi/\sqrt{\bar{n}}$. For reasonable values of $\bar{n}$, this does not make it too difficult to achieve reasonable values of single-photon optomechanical coupling rate.

\subsection{Conditional mechanical state based on photo-detection}
\subsubsection{Without mechanical damping}
In the case of the opto-mechanical model, the conditional state of the mechanics depends on the time of detection, $t_1$, which is itself a random variable. This is quite different to the simple Jaynes-Cummings measurement model considered in the previous section where the interaction time $\tau$  was under the control of the experimenter.  However, in the ideal case, if a photon is emitted from cavity $a_2$ it will be detected at $D_2$. As the emission rate is simply proportional to the probability of there being one photon in this cavity, a detection at some time $t_1$ is equivalent to a direct readout of the qubit state $|1\rangle$ in the Jaynes-Cummings model with an interaction time $t_1$ although, unlike that model, the qubit is destroyed in the process. 

However there is an additional new feature. A photon detection at $D_2$ can occur in two indistinguishable ways: the photon can be reflected from cavity $a_2$ directly into the detector or it can be transmitted from within the cavity after first being absorbed. This feature is reflected in the jump operator $J$ as the sum of two terms, $J=\sqrt{\gamma}c+\sqrt{\kappa_2}a_2$. This can lead to an interference term in the detection rate. 

We also need to include the possibility that a photon is lost from cavity $a_1$. As there is at most one photon in the entire system at any time, a photon lost from $a_1$ means that no photon can be counted at $D_2$. Thus loss from the other cavity appears as a non unit efficiency in the detection process. One could, of course, insert a detector at the output from $a_1$ to herald such erroneous events.  If we do not monitor this channel, we will need to define a cut-off time: non-detection at $D_2$ up to the cut-off time indicates failure and we simply discard that run and start again. 

 We first  compute the conditional state conditioned on no detections, either at $D_1$ or $D_2$,  up to time $t$. The quantum theory of continuous photon counting \cite{SD} shows that this (unnormalised) conditional state is determined by 
\begin{equation}
\tilde{\rho}^{(0,0)}(t)={\cal S}(t)\rho(0) ,
\end{equation}
where the superscript is defined by $(n_1,n_2)$ where $n_i$ is the count number recorded at detector $D_i$. 
where ${\cal S}(t)\rho(0)=\tilde{\rho}^{(0,0)}(t)$ is the conditional state given no counts up to time $t$ and is given by solving
\begin{equation}
\label{no-count}
\frac{d\tilde{\rho}}{dt}=-i(K\tilde{\rho}-\tilde{\rho} K^\dagger) ,
\end{equation}
where the non-Hermitian operator $K$ is given by
\begin{equation}
\label{K}
K=g(b^\dagger a_1^\dagger a_2+ba_1 a_2^\dagger)  -i\sqrt{\gamma\kappa_2}(ca_2^\dagger-c^\dagger a_2)/2-iJ^\dagger J/2-i\kappa_1 a^\dagger_1 a_1/2 .
\end{equation}
The normalisation of this state is simply the probability for no counts up to time $t$,
\begin{equation}
p(n_1=0,n_2=0,t)={\rm tr}[\tilde{\rho}^{(0,0)}(t)] .
\end{equation}
\begin{figure}[h!]
   \centering
   \includegraphics{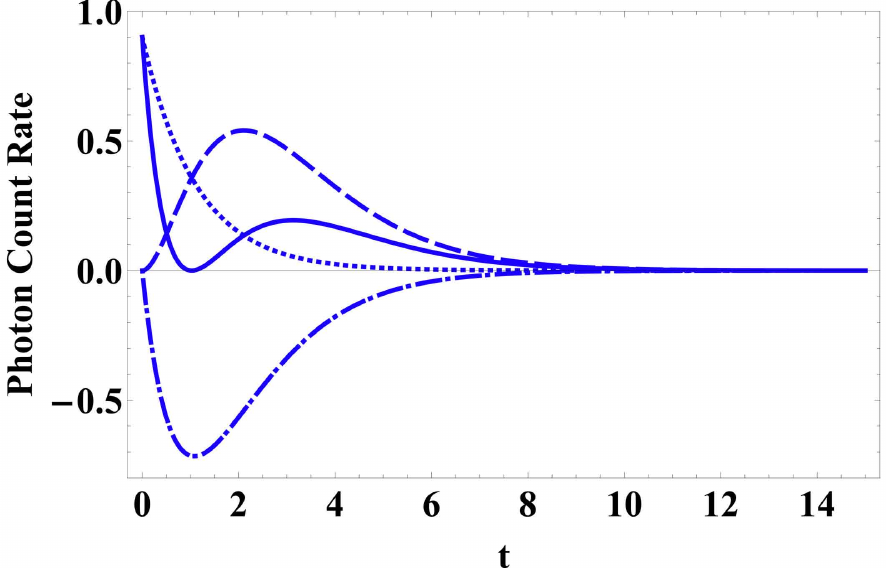} 
   \caption{The photon detection rate (solid line) as the sum of three terms: source emission rate (dotted line), cavity emission rate (dashed line), interference emission rate (dashed-dotted line) showing how interference leads to a minimum at finite time. The zero is due to interference between the two indistinguishable ways a single photon can be detected: reflected from the input mirror or absorbed by the cavity and subsequently re-emitted. The parameters are $g=0$ and $\kappa_1 =0$. }
   \label{fig5}
\end{figure} 
Note, that if the initial state is pure, we need only to solve the effective Schr\"{o}dinger equation
\begin{equation}
\label{Sch}
\frac{d|\tilde{\psi}^{(0)}\rangle }{dt}= -iK|\tilde{\psi}^{(0)}\rangle ,
\end{equation}
to give
\begin{equation}
\tilde{\rho}^{(0,0)}(t)=|\tilde{\psi}^{(0}(t)\rangle\langle \tilde{\psi}^{(0)}(t)| .
\end{equation}

We now ask for the conditional state of the system given that one photon is counted at $D_2$ in time $t$ to $t+dt$. Such an event means that no photon can have been decayed through the output of cavity $a_1$.  This conditional  state is 
\begin{equation}
\tilde{\rho}^{(0,1)}(t)=J{\cal S}(t)\rho(0)J^\dagger .
\end{equation}
If the initial state is a pure state, this conditional state is also a pure state
\begin{equation}
|\tilde{\psi}^{(0,1)}\rangle =J|\tilde{\psi}^{(0)}(t)\rangle =\sqrt{\gamma}c|\tilde{\psi}^{(0)}(t)\rangle +\sqrt{\kappa_2}a_2|\tilde{\psi}^{(0)}(t)\rangle ,
\end{equation}
which is superposition of the two ways in which a photon can be counted: direct reflection form the cavity or emission from inside the cavity.  This leads to an interference term in the detection rate
\begin{eqnarray}
R_{01}(t) & = &  \gamma \langle \tilde{\psi}^{(0)}(t)|c^\dagger c|\tilde{\psi}^{(0)}(t)\rangle+\kappa_2 \langle \tilde{\psi}^{(0)}(t)|a_2^\dagger a_2|\tilde{\psi}^{(0}(t)\rangle\\\nonumber
& & \ \ \ \  \ \ + \sqrt{\gamma\kappa_2}\left ( \langle \tilde{\psi}^{(0)}(t)|c^\dagger a_2|\tilde{\psi}^{(0}(t)\rangle+c.c\right ) .
\end{eqnarray}
For example, if $\kappa_1=0$ and $g=0$, we find that
\begin{equation}
\label{rate}
R_{01}(t)=\gamma e^{-\gamma t}+\kappa_2 n_2(t)-\frac{4\gamma\kappa_2}{\kappa_2-\gamma}e^{-\gamma t/2}\left (e^{-\gamma t/2}-e^{-\kappa_2 t/2}\right ),
\end{equation}
where the mean photon number in cavity $a_2$ is 
\begin{equation}
n_2(t)=\frac{4\gamma\kappa_2}{(\kappa_2-\gamma)^2}\left (e^{-\gamma t/2}-e^{-\kappa_2 t/2}\right )^2 .
\end{equation}
The first two terms in Eq. (\ref{rate}) correspond to direct detection from the source cavity and $a_2$ respectively. The last term represents the interference between these two indistinguishable events and leads to a zero in the detection rate at $D_2$ as a function of time (figure 5).

If the photon decays through cavity $a_1$, it can never be detected at $D_2$. However if we do not monitor this output we have no way of knowing when this happens. We thus need to sum over all times at which a photon could be emitted from cavity $a_1$.  Note also that once a photon is lost the operation ${\cal S}$ acts trivially as the identity. The (unnormalised) conditional state given that one photon was lost from cavity $a_1$ at any time over the interval $[0,t)$ is  
\begin{equation}
\tilde{\rho}^{(1,0)}([0,t))=\kappa_1 \int_0^t dt_1 a_1 e^{-iK t_1}\rho(0)e^{K^\dagger t_1} a_1^\dagger .
\end{equation}
 The normalisation of this state is the error probability 
 \begin{equation}
 p_{err}(t)= \kappa_1 \int_0^t dt_1 {\rm tr}\left  [a_1 e^{-iK t_1}\rho(0)e^{K^\dagger t_1} a_1^\dagger\right ] ,
 \end{equation}
  as this represents the probability that a photon was lost from the system before it could be detected at $D_2$. Clearly to keep the error low we need to ensure $\kappa_1 <<\kappa_2$. 
  
  If we expand the initial mechanical state in the eigenstates of $b^\dagger b$ which we write as $|n\rangle_b$ the dynamics is closed in the three dimensional subspace spanned by the basis 
\begin{equation}
\{|1\rangle_n,|2\rangle_n,|3\rangle_n\}=\{|1,0,0,n\rangle, |0,0,1,n\rangle, |0,1,0,n+1\rangle\},
\end{equation}
where $|x,y,z,n\rangle\equiv |x\rangle_c|y\rangle_1|z\rangle_2|n\rangle_b$. We can now expand
\begin{equation}
\label{Schsoln}
|\tilde{\psi}^{(0)}(t)\rangle=\sum_{k=1}^3 c_k^n(t)|k\rangle_n .
\end{equation}
Substituting this into Eq.(\ref{Sch}) gives a closed set of equations for the coefficients that can be solved analytically. For future purposes we rewrite the initial state in Eq.(\ref{initial}) in another form
\begin{equation}
|\Psi(0)\rangle=\sum_{n=0}^\infty \beta_n(t_0)|1\rangle_n ,
\end{equation}
 where $\beta_n(t_0)=e^{-\beta^2/2}\frac{\beta^n}{\sqrt{n!}}$. Therefore, the initial number distribution is the Poisson distribution
 \begin{equation}
 P_n^0(t_0)=e^{-|\beta|^2/2} \frac{|\beta|^{2n}}{n!} .
 \end{equation}

As discussed in section \ref{simple-JC-model}, we need to be in strong opto-mechanical coupling regime for which $g$ is of the order of $\kappa_2$ in order to reach (or get close to) a Fock state within a reasonable number of trials.  Hence, in units such that $\kappa_{2}=1$, we have $g=1$. We also assume that $\kappa_{1}=0.2$, $\gamma=0.9$ and the mechanical damping rate, $\gamma_m<<\kappa_2$, so that we can neglect the mechanical damping. We consider mechanical damping and thermal effects in subsection \ref{mech-dmap}.

Figure 6 shows the detection rate for the first measurement for the given parameters. 
\begin{figure}[h!]
   \centering
   \includegraphics{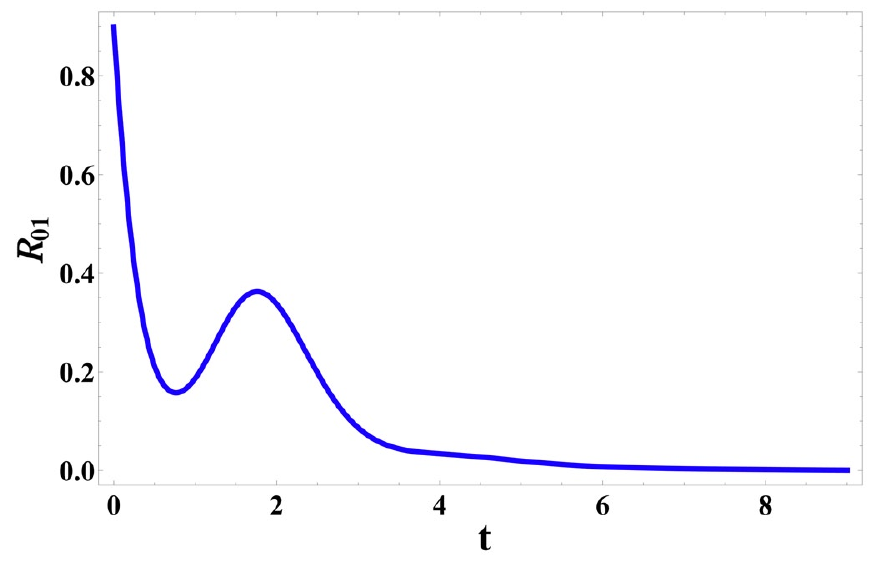} 
   \caption{The photon detection rate for $\beta=2$, $g=1$, $\kappa_2 =1$ and $\kappa_1=0.2$. }
   \label{Fig6}
\end{figure}  
We use the rate function to generate a random detection time $t_1$ in a time interval which starts just after the minimum and ends close where the detection rate is nearly zero.  This choice assures us that, with high probability,  we are detecting a photon from cavity $a_2$, after it has interacted with the mechanical system not one which is reflected of the mirror directly from the source. We then substitute $t_1$ in Eq.(\ref{Schsoln}) to get the normalised conditional state of the system given no counts up to $t_1$
\begin{equation}
|\psi^{(0)}(t_1)\rangle=\frac{|\tilde{\psi}^{(0)}(t_1)\rangle}{\sqrt{\langle\tilde{\psi}^{(0)}(t_1)|\tilde{\psi}^{(0)}(t_1)\rangle}} .
\end{equation}  
Applying jump operator $J$ on $|\psi^{(0)}(t_1)\rangle$ we get the conditional state of the system $|\tilde{\psi}^{(1)}(t_1)\rangle$ given that one photon is counted in time $t_1$ to $t_1+dt_1$. Normalising this state we get
\begin{equation}
|\psi^{(1)}(t_1)\rangle=\sum_{n=0}^\infty \beta_n(t_1)|0\rangle_n ,
\end{equation}
where
\begin{equation}
\beta_n(t_1)=\frac{\beta_n(t_0)(\sqrt{\gamma}c_1^n(t_1)+\sqrt{\kappa_2}c_2^n(t_1))}{\sqrt{\sum_{n=0}^\infty \beta_n^2 (t_0)|\sqrt{\gamma}c_1^n(t_1)+\sqrt{\kappa_2}c_2^n(t_1)|^2}} .
\end{equation}
The phonon number distribution for the conditional state now becomes
\begin{equation}
P_n^1(t_1)=P_n^0(t_0)P(n,t_1) ,
\end{equation}
where
\begin{equation}
P(n,t_1)=\frac{|\sqrt{\gamma}c_1^n(t_1)+\sqrt{\kappa_2}c_2^n(t_1)|^2}{\sum_{n=0}^\infty \beta_n^2 (t_0)|\sqrt{\gamma}c_1^n(t_1)+\sqrt{\kappa_2}c_2^n(t_1)|^2} .
\end{equation}
We will repeat the measurement process preparing another single photon in the source and using the new initial state $\sum_{n=0}^\infty \beta_n (t_1)|1\rangle$. Given the state of the system after each measurement, we can re-calculate the detection rate from which we again sample a random detection time.

 After $r$ detection events, the conditional state of the system is
\begin{equation}
\label{cs}
|\psi^{(r)}(t_r)\rangle=\sum_{n=0}^\infty \beta_n(t_r)|0\rangle_n ,
\end{equation}

\begin{figure}[h!]
   \centering
   \includegraphics{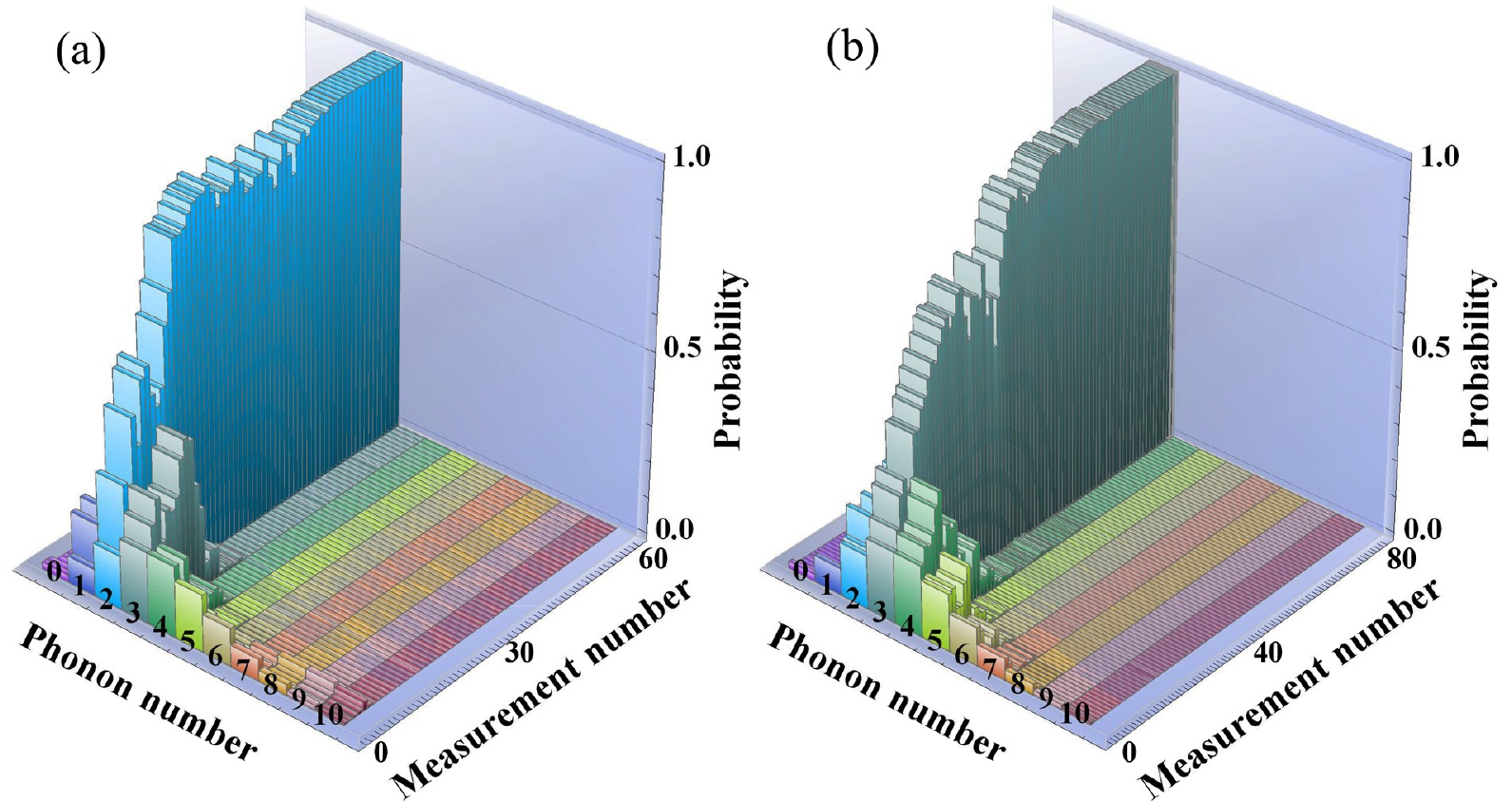} 
   \caption{Phonon number distribution function histograms after successive measurements for $\beta=2$ ($\bar{n}=4$), $g=1$, $\gamma=0.9$, $\kappa_2 =1$, $\kappa_1=0.2$ and random detection times. As we increase the number of measurements, the phonon number distribution evolves from a poissonian distribution into number state distributions (a) n=2 and (b) n=3. }
   \label{fig7}
\end{figure}

where
\begin{equation}
\beta_n(t_r)=\frac{\beta_n(t_{r-1})(\sqrt{\gamma}c_1^n(t_r)+\sqrt{\kappa_2}c_2^n(t_r))}{\sqrt{\sum_{n=0}^\infty \beta_n^2 (t_{r-1})|\sqrt{\gamma}c_1^n(t_r)+\sqrt{\kappa_2}c_2^n(t_r)|^2}} .
\end{equation}
The phonon number distribution function after $r$ measurements now becomes
\begin{equation}
\label{ND}
P_n^r (t_r)=P_n^{r-1} (t_{r-1}) P(n,t_r) ,
\end{equation}
where
\begin{equation}
P(n,t_r)=\frac{|\sqrt{\gamma}c_1^n(t_r)+\sqrt{\kappa_2}c_2^n(t_r)|^2}{\sum_{n=0}^\infty \beta_n^2 (t_{r-1})|\sqrt{\gamma}c_1^n(t_r)+\sqrt{\kappa_2}c_2^n(t_r)|^2} .
\end{equation}

Each measurement provides partial information about phonon number (typically less than one but per trial).  The procedure can be explained looking at Eq.(\ref{ND}) and is quite similar to the simple model with deterministic interaction times discussed in section \ref{simple-JC-model}.  After the $r$'th detection event, the phonon number distribution is multiplied by a filter function $P(n,t_r)$ which, for appropriate values of $t_r$, suppresses certain values of $n$. Continuing the measurement process we can get more information leading to a gradual collapse of the distribution onto a single number state. Figures 7(a) and 7(b) show the phonon number distributions for sixty successive measurements and eighty measurements respectively. The number of measurements required to reach a Fock state  is at least roughly consistent with the prediction of the simple model in  \ref{simple-JC-model}, i.e. scaling as $\bar{n}^2$, despite the stochastic fluctuations in the interaction times in the optomechanical model. 

\begin{figure}[h!]
   \centering
   \includegraphics{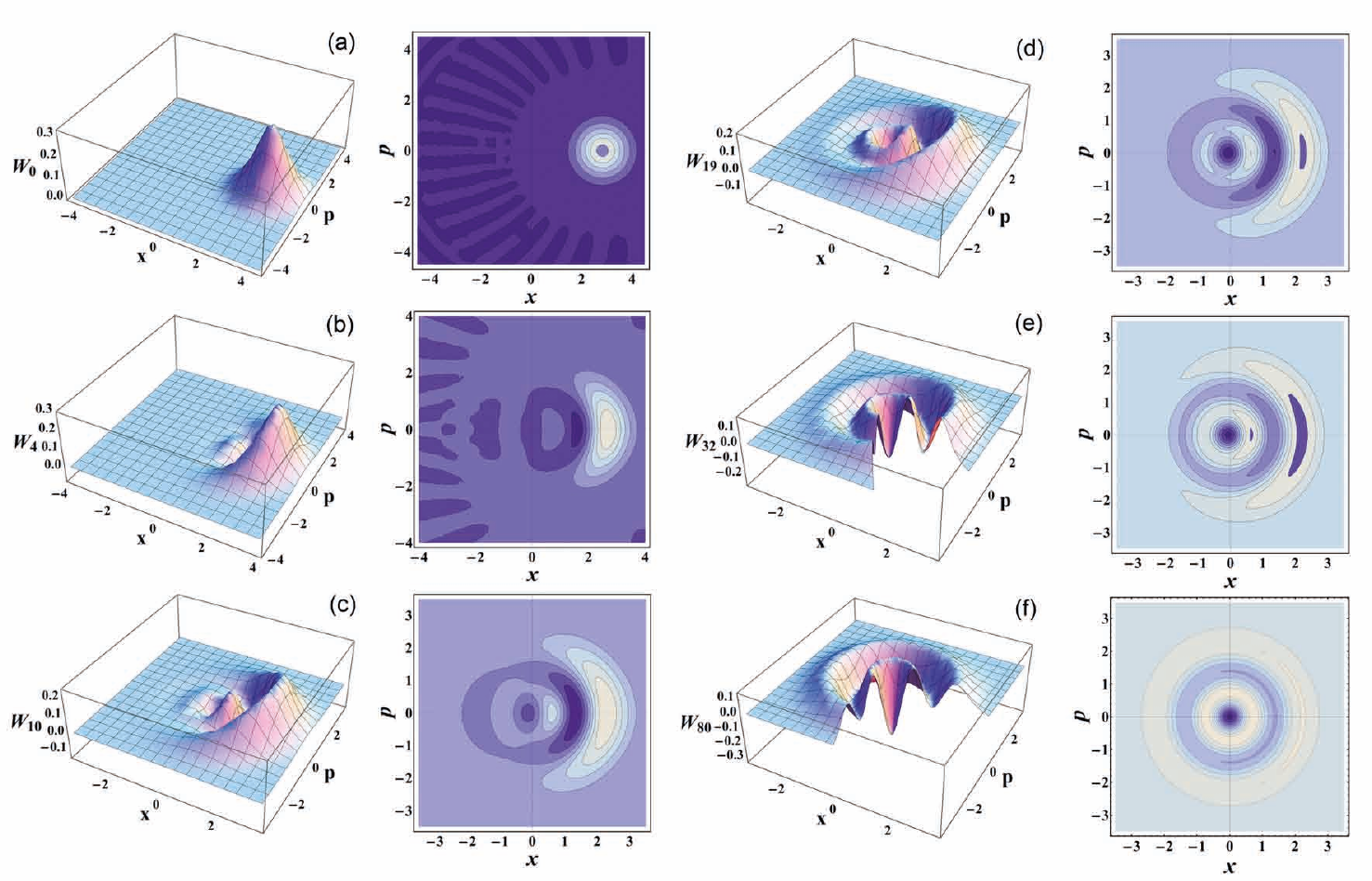} 
   \caption{The evolution of the Wigner distribution function $W_r(x,p)$ with successive measurements where $r$ is the measurement number. $\beta=2$, $g=1$, $\gamma=0.9$, $\kappa_2 =1$, $\kappa_1=0.2$ and the same detection times as in figure 7(b). Measuring the qubit repeatedly we drive the mechanics from a state with a distribution function from a Poisson distribution with mean $\bar{n}=4$ to a  number eigenstate with exact excitation number of $n=3$.  }
   \label{fig8}
\end{figure}

The figure shows the evolution of the number distribution from Poisson distribution into Fock state at $n=2$ and $n=3$  for different choices of the simulated detection times. Once the system collapses to a specific phonon number state, it will remain in that state upon further measurements. (Given that we have neglected mechanical dissipation and thermal fluctuations).  
Some simulations do not settle down to a single number  state but show jumping between two number states so that in some cases we have a competition between two $n$ values in number distribution.  

We can illustrate that the steady state conditional states do indeed tend to number states by computing the  Wigner function.  Figure 8 shows the Wigner functions $W_r(x,p)$ for some arbitrary chosen measurement numbers $r$. The evolution from a Poisson distribution in the number basis to the phase space distribution for the corresponding number state with $n=3$ is clear. As the phonon number converges, the phase uncertainty increases until we get full information about phonon number and phase uncertainty becomes complete.

\subsubsection{With mechanical damping}
\label{mech-dmap}

Our simulations show a collapse to a mechanical phonon number state is possible even including loss from the second cavity. In an experiment the measurement induced collapse will be competing with thermal fluctuations and dissipation. The measurement induced collapse proceeds at a rate determined by the count rate which is bounded by  $\kappa$.  We can estimate the rate of change of the phonon number variance due to the damping of the mechanics if we describe the mechanical dissipation using the usual weak damping master equation of quantum optics,
\begin{equation}
\label{thermal_me}
\frac{d\rho_m}{dt}= {\cal L}\rho+\gamma_m(\bar{N}+1){\cal D}[b]\rho_m+\gamma_m\bar{N}{\cal D}[b^\dagger]\rho_m
\end{equation}
where the first term is the opto-mechanical part and  $\gamma_m$ is the mechanical damping rate and $\bar{N}$ is the mean thermal phonon excitation number at the mechanical resonance frequency.  This equation indicates that phonons enter the mechanical resonator at a rate determined by $\gamma_m\bar{N}$ and decay at a rate determined by $\gamma_m(\bar{N}+1)$. The overall effect is a rate of increase for the number variance at a rate proportional to $\gamma_m\bar{N}$.  Thus the measurement induced state reduction will dominate the increase due to mechanical dissipation  if $\kappa>>\gamma_m \bar{N}$. We now consider the competition between thermalisation and the reduction in the number fluctuations due to successive measurements. 

\begin{figure}[h!]
   \centering
   \includegraphics{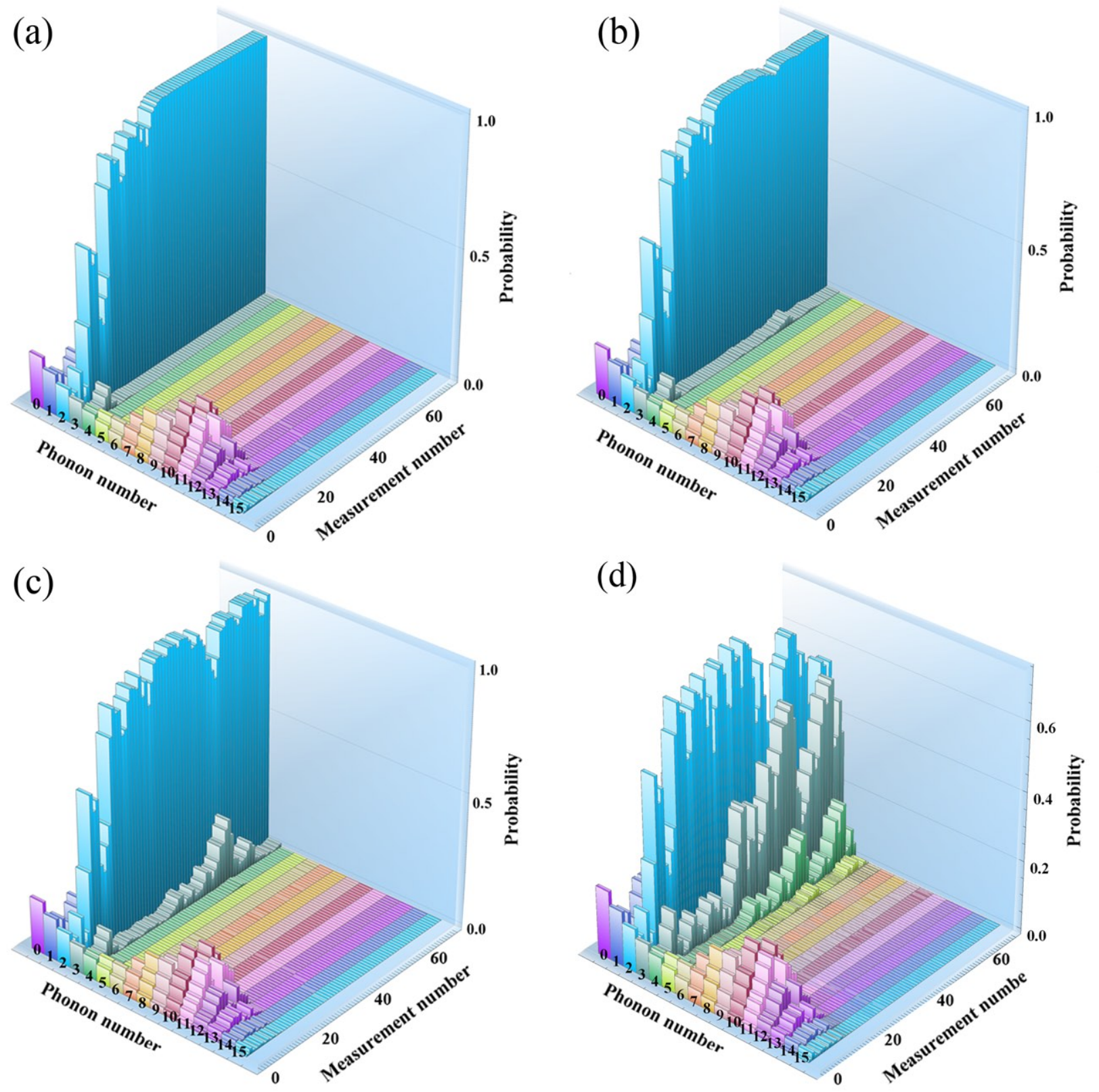} 
   \caption{Phonon number distribution function histograms after successive measurements for $\bar N=4$, $g=1$, $\kappa_2 =1$, $\kappa_1=0.2$, $\gamma=0.9$ and different values of mechanical damping rate: (a) $\gamma_m=0$, (b) $\gamma_m=0.00001$, (c) $\gamma_m=0.0001$, (d) $\gamma_m=0.001$. In (b) we have the optimum value for mechanical damping rate for this scheme, so that we conditionally drive the mechanical resonator to a number state. We see the broadening effect of the mechanical damping rate as we increase $\gamma_m$.}
   \label{fig9}
\end{figure}

If the system is started in the thermal equilibrium state that follows from Eq.(\ref{thermal_me}), the initial state of the mechanical degree of freedom is a thermal state with the number distribution
\begin{equation}
P_n^0 (t_0)=\frac{1}{1+\bar N}\sum_{n=0}^\infty (\frac{\bar N}{1+\bar N})^n.
\end{equation}
The rate at with measurements can proceed is limited by the time taken for each detection, so we will consider the case in which the single photon source is loaded immediately after each detection. We now include the effect of mechanical damping in the conditional Schro\"{o}dinger for no detections up to time $t_r$, 
\begin{equation}
\frac{d\tilde{\rho}}{dt}=-i(K\tilde{\rho}-\tilde{\rho} K^\dagger)+\gamma_m(\bar{N}+1){\cal D}[b]\tilde{\rho}+\gamma_m\bar{N}{\cal D}[b^\dagger]\tilde{\rho} ,
\end{equation}
where $K$ is given by the Eq.(\ref{K}) and the damping superoperators are defined by Eq.(\ref{superoperator}).

Figure (\ref{fig9}) shows the phonon number distributions after each readout for seventy measurements. When the mechanical damping rate is zero (figure 9(a)), the number distribution evolves from a thermal distribution into a $n=2$ peak. If we increase the damping rate gradually, for $\gamma_m/\kappa_2\gtrsim10^{-5}$ we observe that the number distribution starts to broaden. It is clear from this discussion that the rate at which phonons enter and exit the mechanical resonator is the key limiting factor in reaching a Fock state.  Thus the coupling of the mechanical resonator to its environment will need to be carefully engineered to make $\gamma_m\bar{N}$ as small as possible, perhaps using an external phononic bandgap shield\cite{Chan}, or carefully engineered supports\cite{Cole}.  In the case of \cite{Painter} a continuous-flow helium cryostat provides pre-cooling  to achieve a bath occupancy of the a $3.68$ GHz mechanical mode of $\bar{N}<100$.

\section{Conclusion}
We have considered a quantum opto-mechanical system based on the coupling between a dual rail optical qubit code, formed of two optical cavity modes, and a single bosonic matter degree of freedom in the strong coupling limit for which the single photon opto-mechanical coupling rate $g$ is of the order of the decay rate of the cavity $\kappa$ and shown how single photon detection can conditionally drive the mechanical resonator to a phonon number state provided the mechanical thermalisation rate $\gamma_m\bar{N}$ is small enough.  Despite  the fact that in this realisation the interaction time between the mechanics and dual rail  optical qubit is a random variable, we can found regimes in which the mechanical phonon number distributions become sharply peaked at a particular value of $n$. Successive photon counts, even though randomly distributed,  provide a  record of the interaction time in each measurement which is sufficient to  gain information about the state of the excitation number of the mechanics in a sequence of detection events. 

The achievement of the strong coupling regime will be a challenge.   Several other experimental groups are working towards achieving strong coupling at the single photon level, \cite{murch,eichenfield,purdy,roh,ding1,ding2,gavartin}.  The required values for the mechanical damping rate are achievable, for example, Chan et al. \cite{Chan}, have  $\gamma_{m}=7.5$kHz for $\kappa/2\pi=214$MHz, $g=1.1$MHz and $\omega_m=5.1$GHz.  In the dimensionless units of this work these are, $\gamma_{m}/\kappa=3.5\times 10^{-5}$ for $g/\kappa= 5.1 \times 10^{-3}$ and $\omega_m/\kappa=30$.
The required value for the thermalisation rate  $\gamma_m\bar{N}$ will be a challenge.  et al. \cite{Kippenberg} using a cryostat at $0.65$K have $\gamma_m\bar{N}/2\pi = 2$MHz which with a optical decay rate of $\kappa=10$MHz corresponds to a dimensionless thermalisation rate of $0.3$. We expect continued advances in the design and fabrication of  electromechanical and optomechanical systems will enable sufficient isolation from the thermal environment to reach a regime where the measurement induced collapse of the phonon distribution can beat the broadening due to thermal effects.

\subsection*{Acknowledgments}
S. Basiri-Esfahani acknowledges support from the University of Queensland International Scholarship (UQI). S. Basiri-Esfahani also acknowledges discussions with S. Rahimi-Keshari.  This work was supported by the Australian Research Council Centre of Excellence  for Engineered Quantum Systems grant number CE110001013.

\end{document}